\documentclass[12pt,preprint]{aastex}
\begin{document}
\title{Modeling the Anisotropic Tidal Effect on the Spin-Spin Correlations of Low-Mass Galactic Halos}
\author{Jounghun Lee}
\affil{Astronomy Program, Department of Physics and Astronomy, Seoul National University, 
Seoul 08826, Republic of Korea} 
\email{jounghun@astro.snu.ac.kr}
\newcommand{\etal}{{\it et al.}}
\newcommand{\beq}{\begin{equation}}
\newcommand{\eeq}{\end{equation}}
\newcommand{\ben}{\begin{eqnarray}}
\newcommand{\een}{\end{eqnarray}}
\newcommand{\hbs}{\hat{\bf s}}
\newcommand{\hbsp}{\hat{\bf s}^{\prime}}
\newcommand{\hs}{\hat{s}}
\newcommand{\hT}{\hat{T}}
\newcommand{\pT}{T^{\prime}}
\newcommand{\hbT}{\hat{\bf T}}
\newcommand{\tbT}{\tilde{\bf T}}
\newcommand{\hTp}{\hat{T}^{\prime}}
\newcommand{\hbTp}{\hat{\bf T}^{\prime}}
\newcommand{\bT}{{\bf T}}
\newcommand{\tT}{\tilde{T}}
\newcommand{\tTp}{\tilde{T}^{\prime}}
\newcommand{\tJ}{\tilde{J}}
\newcommand{\hk}{\hat{k}}
\newcommand{\hr}{\hat{r}}
\newcommand{\bk}{{\bf k}}
\newcommand{\by}{{\bf y}}
\newcommand{\bx}{{\bf x}}
\newcommand{\bxp}{{\bf x}^{\prime}}
\newcommand{\bt}{{\bf t}}
\newcommand{\hby}{\hat{\bf y}}
\newcommand{\hbd}{\hat{\bf d}}
\newcommand{\hbx}{\hat{\bf x}}
\newcommand{\hbr}{\hat{\bf r}}
\newcommand{\hbt}{\hat{\bf t}}
\newcommand{\br}{{\bf r}}
\newcommand{\bfv}{{\bf v}}
\newcommand{\bq}{{\bf q}}
\newcommand{\bI}{{\bf I}}
\newcommand{\txi}{\tilde{\xi}}
\newcommand{\lam}{\lambda}
\newcommand{\hlam}{\hat{\lambda}}
\newcommand{\hbp}{\hat{\bf p}}
\newcommand{\sig}{\sigma}

\begin{abstract}
The halo spin-spin correlation function, $\eta(r)$, measures how rapidly the strength of the alignments of the spin directions between 
the neighbor halos change with the separation distance, $r$.  The previous model based on the tidal torque theory expresses the halo 
spin-spin correlation function as a power of the linear density two-point correlation function, $\eta(r)\propto \xi^{n}(r)$, predicting $n=2$ 
in the linear regime and $n=1$ in the non-linear regime. 
Using a high-resolution N-body simulation, we show that the halo spin-spin correlation function in fact drops much less rapidly with 
$r$ than the prediction of the previous model, finding $\eta(r)$ to be statistically significant even at $r\ge 10\,h^{-1}$Mpc on the dwarf 
galaxy scale. 
Claiming that the anisotropic tidal effect is responsible for the failure of the previous model, we propose a new formula for the halo 
spin-spin correlation function expressed in terms of the integrals of $\xi(r)$. The new formula with the best-fit parameters turns out to 
agree excellently with the numerical results in a broad mass range, $0.05\le M/(10^{11}\,h^{-1}\,M_{\odot})\le 50$, 
describing well the large-scale tail of $\eta(r)$. We discuss a possibility of using the large-scale spin-spin correlations of the dwarf galactic 
halos as a complementary probe of dark matter. 
\end{abstract}
\keywords{cosmology:theory --- large-scale structure of universe}
\section{Introduction}\label{sec:intro}

One of the main missions of modern cosmology is to determine the initial conditions of the universe from the observables. For the completion 
of this mission, such linear observables as the cosmic microwave background radiation (CMB), large scale velocity flows, baryonic acoustic 
oscillations (BAO) and etc., which do not evolve much from the initial states and thus can be well described by the first order perturbation theory, 
have been regarded as the most optimal diagnostics \citep[e.g.,][]{vit-etal86,boom00,SE03}. 
Powerful as they are as a probe of cosmology, the simultaneous dependence of these linear observables on the multiple parameters that are required to 
describe the cosmological initial conditions often invokes the degeneracy problem. 
For example, it was recently shown by \citet{PR18} that although the Planck satellite experiment advocates a flat $\Lambda$CDM universe 
whose energy density is dominated by the cosmological constant $\Lambda$ and cold dark matter (CDM) with zero spatial curvature \citep{planck13},  
a non-flat $\Lambda$CDM universe fits the CMB and BAO data on the large-scale equally well if the Hubble constant, $H_{0}$, has a much lower 
value than the Planck best-fit result.

Prospecting the near-field non-linear observables for complementary cosmological probes has been on the rise to overcome the limitations of the linear 
counterparts, in spite of their complicated nature that often defies any analytical approaches \citep{BP06}.  
Given the high-energy physics often leave imprints on the small-scales \citep[for a review, see][]{bia19}, 
it was suggested that a prominent diagnostics based on the non-linear observables, if found, should enable us not only to break the parameter degeneracy 
but also to open a new window on the early universe.
Among various nonlinear observables suggested so far as probes of cosmology such as the compact mini-halos, dynamics of the Local Group, 
wide binary stars, velocity distribution function of galaxy clusters, properties of neutron stars, and so on 
\citep[e.g.,][]{asl-etal16,car-etal17,nta-etal17,BZ18,SY19}, 
the galaxy spin-spin correlation function has recently garnered astute attentions because of its good prospects for the practical application. 
For instance, \citet{sch-etal15} claimed that the anisotropic inflation models can be tested and constrained by measuring the galaxy shape-shape 
(or spin-spin) correlation function \citep[see also][]{chi-etal16,kog-etal18}.  Very recently, \citet{yu-etal19} found it possible in principle to detect a signal of 
the spontaneous breaking of chiral symmetry predicted by the quantum chromodynamics from the measurement of the galaxy spin-spin correlation function.

Constructing a solid theoretical framework for the galaxy spin-spin correlation function is a prerequisite toward its success as a probe of cosmology. 
It was \citet{pen-etal00} who for the first time developed an analytic formula for the galaxy spin-spin correlation function, $\eta(r)$, based on the linear tidal 
torque theory \citep{dor70,whi84}, according to which $\eta(r)$ is proportional to the square of the linear density two-point correlation function, $\xi^{2}(r)$.  
Their model, however, turned out to fail in matching on a quantitative level the numerical results from N-body simulations in which $\eta(r)$ was 
found to drop with $r$ not so rapidly as $\xi^{2}(r)$.  
Ascribing this disagreement to the development of the non-Gaussianity of the tidal fields in the non-linear regime,  \citet{HZ02} claimed that 
$\eta(r)$ in the nonlinear regime should be described as a linear scaling of $\xi(r)$ \citep[see also][]{HZ08}. 
Their claim of $\eta(r)\propto\xi(r)$ was later confirmed by \citet{lp08} at low-redshifts ($z\le 0.5$) in the halo mass range of 
$10^{11}\le M/(h^{-1}\,M_{\odot})\le 10^{13}$. 

Although the linear scaling of $\eta(r)$ with $\xi(r)$ was found to work quite well at distances of  $r<10\,h^{-1}$Mpc, it turned out to fail in describing 
the tail of $\eta(r)$ at larger distances $r\ge 10\,h^{-1}$Mpc \citep{lp08}. Moreover, this model has an conceptual downside: the anisotropic tidal effect has not been 
properly taken into account, which is likely to be the cause of its failure at $r\ge 10\,h^{-1}$Mpc.  Among many aspects of the evolved tidal fields, it should not be 
only the non-Gaussianity that leads the spin-spin correlation function to decrease slowly with $r$.  As a matter of fact, the growth of the anisotropy in the evolved 
tidal fields may contribute even more to the generation of the galaxy spin-spin correlation at large scales, given the recent numerical finding that the cosmic web 
generated by the anisotropic tidal fields are closely linked with the intrinsic spin alignments of the low-mass galaxies \citep{cod-etal15a}.  In this Paper, we attempt to 
find a new improved formula for $\eta(r)$ that is valid at larger distances even on the dwarf galaxy scale by taking the anisotropic tidal effects into consideration. 
Our analysis will be done in the framework of the extended model for the tidally induced spin alignments recently proposed by \citet{lee19} to describe the 
intrinsic alignments between the directions of the galaxy spins (and shapes) and the eigenvectors of the local tidal tensors. 

The outlines of the upcoming Sections are as follows. Section \ref{sec:review_etm} is devoted to reviewing the extended model for the tidally induced spin 
alignments on which a new formula for $\eta(r)$ will be based.  Section \ref{sec:etm_extension} is spared to prove the validity of the extended model for the 
tidally induced spin alignment on the dwarf galaxy scale. Section \ref{sec:review_eta} is devoted to reviewing the previous models for the spin-spin correlation 
functions and to explaining their limitations as well as their merits. Section \ref{sec:eta_ani} presents a new formula for the galaxy spin-spin correlation function 
based on the extended model for the tidally induced spin alignments. Section \ref{sec:eta_test} is spared to show how successfully the new formula for the galaxy 
spin-spin correlation function survives a numerical test in a broad mass range at various redshifts. The summary of our achievements and the discussion of the 
future application of our new model are presented in Section \ref{sec:sum}.

\section{Tidally Induced Spin Alignments on the Dwarf Galaxy Scales}

\subsection{Review of the Analytic Framework}\label{sec:review_etm}

Throughout this Paper, we will let the unit spin vector of a dark matter halo, unit traceless tidal tensor surrounding a halo, 
set of three eigenvalues of the unit traceless tidal tensor in a decreasing order and set of the corresponding tidal eigenvectors be 
denoted by $\hbs=(\hs_{i})$, $\hbT=(\hT_{ij})$, $\{\hlam_{1}, \hlam_{2},\hlam_{3}\}$ and $\{\hbp_{1},\hbp_{2},\hbp_{3}\}$, 
respectively. We will also let $s$, $M$ and $R_{f}$ denote the magnitude of the spin vector, halo mass, and 
smoothing scale, respectively.  

The original model developed by \citet{lp00} for the tidally induced spin alignments of dark matter halos assumes that the 
conditional probability density function, $p(s,\hbs\vert\hbT)$, follows a multi-variate Gaussian distribution, and that the conditional 
covariance, $\langle\hs_{i}\hs_{j}\vert{\bf T}\rangle$, can be expressed in terms of the anti-symmetric product of $\hbT$ as 
\beq
\label{eqn:model1}
\langle\hs_{i}\hs_{j}\vert\hbT\rangle =  \left(\frac{1}{3}+\frac{3}{5}c_{t}\right)\delta_{ij} - \frac{3}{5}c_{t}\hT_{ik}\hT_{kj}\,  ,
\eeq
where $c_{t}$, called the spin correlation parameter, ranges from $0$ to $1$. Equation (\ref{eqn:model1}) provides the simplest 
description of the $\hat{\bf s}$-$\hat{\bf p}_{2}$ alignment whose presence was naturally predicted by the linear tidal torque theory 
\citep{dor70,whi84} but whose strength cannot be determined from the first principle due to its stochastic nature \citep{lp00,por-etal02}. 

Rearranging the terms in Equation (\ref{eqn:model1}) about $c_{t}$ in the principal frame of $\hbT$ gives 
\beq
\label{eqn:ct}
c_{t}=10\left(\frac{1}{3}-\sum_{i=1}^{3}\hlam^{2}_{i}\hs^{2}_{i}\right)\, ,
\eeq
which translates a larger value of $c_{t}$ into a stronger $\hat{\bf s}$-$\hat{\bf p}_{2}$ alignment. Before the turn-around moment when 
$(s,\hbs)$ continues to grow under the influence of $\hbT$, $c_{t}$ will increase with time.  At the turn-around moment when $c_{t}$ reaches 
the maximum value of unity, the tidal interaction between $\hbT$ and $\hbs$ will be terminated. After the turn-around moment when 
both of $\hbT$ and $\hbs$ grow nonlinearly, being decoupled from each other, $c_{t}$ would gradually diminish from unity. 
Therefore, the spin correlation parameter, $c_{t}$, is expected to depend on $M$ and $z$, having lower values for the case 
of lower-mass halos at lower redshifts, which must have turned around at earlier epochs. 

Multiple N-body simulations limited the validity of Equation (\ref{eqn:model1}) to $M\ge M_{c}$ with $M_{c}\approx 5\times10^{12}\,h^{-1}\,M_{\odot}$ 
\citep{ara-etal07,hah-etal07,paz-etal08,zha-etal09,cod-etal12,lib-etal13,tro-etal13,vee-etal18}, demonstrating that the halos with $M< M_{c}$ exhibit the 
$\hbs$-$\hbp_{3}$ rather than $\hbs$-$\hbp_{2}$ alignment. Although much effort was made to explain this {\it spin flip} phenomena, 
it has yet to be fully understood what the physical meaning of $M_{c}$ is 
\citep{BF12,LP12,cod-etal12,lib-etal13,wel-etal14,cod-etal15b,lai-etal15,BF16,vee-etal18}. 

In line with this effort, \citet{lee19} recently put forth an {\it extended model for the tidally induced spin alignments} by adding to Equation (\ref{eqn:model1}) 
a new term proportional to $\hbT$ in order to describe the change of the alignment tendency between $M<M_{c}$ and $M\ge M_{c}$.
\beq
\label{eqn:model2}
\langle\hs_{i}\hs_{j}\vert\hbT\rangle =  \left(\frac{1}{3}+\frac{3}{5}c_{t}+\frac{3}{5}d_{t}\right)\delta_{ij} - 
\frac{3}{5}c_{t}\sum_{k=1}^{3}\hat{T}_{ik}\hat{T}_{kj} - \frac{3}{5}d_{t}\hat{T}_{ij}\,  ,
\eeq
where an additional parameter (called the second spin correlation parameter), $d_{t}$, ranging from $0$ to $1$, is introduced to quantify 
the $\hbs$-$\hbp_{3}$ alignment.  

\citet{lee19} derived a formula for $d_{t}$  in terms of $(\hlam_{i})$ and $(\hs_{i})$ from Equation (\ref{eqn:model2}) as 
\beq
\label{eqn:dt}
d_{t} = \frac{3}{5}\sum_{i=1}^{3}\hlam_{i}\hs^{2}_{i}\, ,
\eeq
which translates a larger value of $d_{t}$ into a stronger $\hbs$-$\hbp_{3}$ alignment. It was also shown by \citet{lee19} that 
the value of $d_{t}$ obtained by Equation (\ref{eqn:dt}) increases as $M$ decreases and that the ratio of $d_{t}$ to $c_{t}$ reaches unity 
at a certain mass scale which turns out to be quite close to the critical mass scale for the spin flip phenomenon, $M_{c}$. 

\citet{lee19} derived three probability density functions, $p(\vert\hbs\cdot\hbp_{1}\vert),\ p(\vert\hbs\cdot\hbp_{2}\vert),\ p(\vert\hbs\cdot\hbp_{3}\vert)$, 
based on this model, 
\begin{eqnarray}
\label{eqn:pro_model2}
p(\vert\hat{\bf p}_{i}\cdot\hat{\bf s}\vert)&=& \frac{1}{2\pi}\int_{0}^{2\pi}\,
\left[\prod_{n=1}^{3}\left(1+c_{t}-3c_{t}\hat{\lambda}^{2}_{n}+d_{t}-3d_{t}\hat{\lambda}_{n}\right)\right]^{-\frac{1}{2}}\times \, 
\nonumber \\
&&\left[\sum_{l=1}^{3}\left(\frac{\vert\hbs\cdot\hbp_{l}\vert}
{1+c_{t}-3c_{t}\hat{\lambda}^{2}_{l}+d_{t}-3d_{t}\hat{\lambda}_{l}}\right)\right]^{-\frac{3}{2}}\,d\phi_{jk}\, .
\end{eqnarray}
where $\phi_{jk}$ is the azimuthal angle defined in the plane normal to $\hbp_{i}$, and proved that all of them were in excellent simultaneous agreement 
with the numerical results without resorting to any fitting procedure, provided that $M$ is in the range of $0.5\le M/(10^{11}\,h^{-1}\,M_{\odot})\le 50$  and 
$\hbT$ smoothed on the scales of $R_{f}\ge 5\,h^{-1}$Mpc. Moreover, it was also shown that Equation (\ref{eqn:pro_model2}) naturally predicts the 
$\hbs$-$\hbp_{2}$ alignment for the case of $M\ge M_{c}$, the $\hbs$-$\hbp_{3}$ alignment for the case of $M<M_{c}$ and the $\hbs$-$\hbp_{1}$ 
anti-alignment for both of the cases \citep{lee19}. 

\subsection{Extension to the Dwarf Galaxy Scales}\label{sec:etm_extension}

Now, we would like to investigate whether or not the extended model for the tidally induced spin alignments is also valid 
for the lower-mass halos with $M<5\times 10^{10}\,h^{-1}\,M_{\odot}$ in the highly nonlinear regime. 
For this investigation, we utilize a dataset from the $\nu^{2}$GC-H2 (New Numerical Galaxy Catalog) simulation conducted by 
\citet{ish-etal15} for the Planck cosmology. The linear box size ($L_{\rm box}$), total number of DM particles ($N_{\rm tp}$), mass 
resolution ($m_{\rm dm}$) of the $\nu^{2}$GC-H2 simulations are as follows: $L_{\rm box}=70\,h^{-1}$Mpc, $N_{\rm tp}=2048^{3}$ 
and $m_{\rm dm}=3.44\times10^{6}\,h^{-1}\,M_{\odot}$. The $\nu^{2}$GC-H2 simulation provides the friends-of-friends (FoF) group 
catalog as well as the Rockstar halo catalog at various snapshopts. The former contains information only on the position (${\bf x}$) 
and number of constituent particles ($N_{p}$) of each halo, while additional information on the spin angular 
momentum and virial mass of each halo is available from the latter \citep{rockstar}. 
From the Rockstar catalog, we extract the {\it distinct low-mass halos} with $M\le 5\times 10^{10}\,h^{-1}\,M_{\odot}$ to the exclusion of 
those halos with $N_{p}<300$ whose spin directions are likely contaminated by the poor-resolutions \citep{bet-etal07}. 

Applying the clouds-in-cell methods to the FoF group catalog at $z=0$ without putting a mass-cut, we construct the density 
contrast field on $256^{3}$ grid points.  Then, we obtain the gravitational potential field by applying the inverse Poisson 
equation to the reconstructed density contrast field, convolve it with a Gaussian filter on the scale of $R_{f}=0.5\,h^{-1}$Mpc 
(corresponding to the highly nonlinear regime), and reconstruct the tidal field by numerically calculating the second derivative of the convolved 
gravitational potential field. 
Through the interpolation of the reconstructed tidal field, we determine $\hbT$ at the location of each halo, and then 
find $\{\hlam_{3}\}_{i=1}^{3}$ and $\{\hbp_{i}\}_{i=1}^{3}$ through a similarity transformation of $\hbT$. 
For each halo, we take the absolute value of the projection of $\hbs$ onto the $i$th eigenvector direction, $\vert\hbs\cdot\hbp_{i}\vert$. 
Dividing the range of $[0,\ 1]$ into multiple bins of the same length, we investigate how frequently the measured value of 
$\vert\hbs\cdot\hbp_{i}\vert$ falls into each bin. This frequency divided by the bin length yields the numerical result of the 
probability density, $p(\vert\hbs\cdot\hbp_{i}\vert)$.

\begin{figure}
\begin{center}
\includegraphics[scale=1.0]{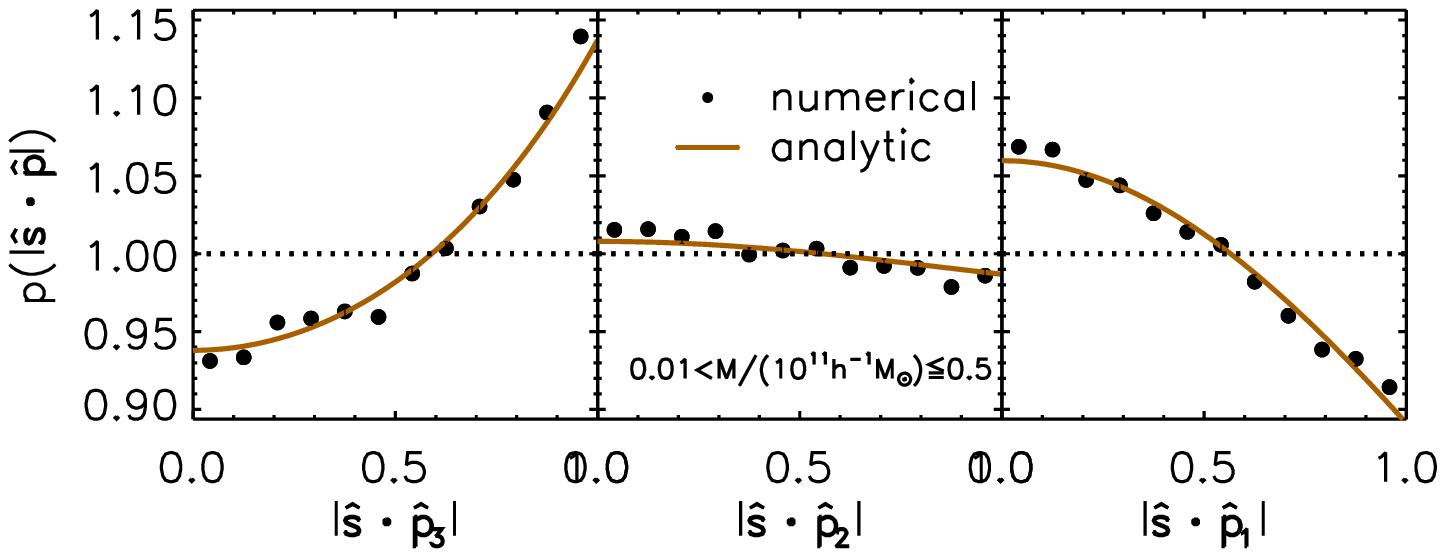}
\caption{Probability density functions of the cosines of the angles of the unit spin vectors of the dwarf  
galactic halos with the third, second and first eigenvectors of the local tidal fields smoothed on the scale 
of $R_{f}=0.5\,h^{-1}$Mpc (left, middle and right panels, respectively).  In each panel, the numerical result 
with errors is displayed as filled circles while the analytic prediction from Equation (\ref{eqn:pro_model2}) 
with $c_{t}=0$ is plotted as brown solid line. The uniform probability density is also plotted as dotted line 
to show the statistical significance of the alignment signals in each panel.}
\label{fig:pro_spin}
\end{center}
\end{figure}

Given the result of \citet{lee19} that the value of $c_{t}$ is negligibly small for $M\le 10^{11}\,h^{-1}\,M_{\odot}$, we set 
$c_{t}=0$ in Equation (\ref{eqn:pro_model2}). Determining $d_{t}$ by Equation (\ref{eqn:dt}) for each halo, we put the mean value of 
$d_{t}$ averaged over the selected halos into Equation (\ref{eqn:pro_model2}) with $c_{t}=0$ to complete the analytic results.  
Figure \ref{fig:pro_spin} compares the analytically determined probability densities (brown solid lines) with the numerical results (filled circles). 
Note the excellent simultaneous agreements between the numerical and analytical results even though no fitting procedure is involved: 
the observed strong $\hbs$-$\hbp_{3}$ alignment, strong $\hbs$-$\hbp_{1}$ anti-alignment and weak $\hbs$-$\hbp_{2}$ anti-alignment are 
simultaneously well described by the analytic results, which confirms the validity of Equations (\ref{eqn:model2})-(\ref{eqn:pro_model2}) in 
the highly nonlinear regime characterized by $M\le 5\times 10^{10}\,h^{-1}\,M_{\odot}$ and $R_{f}=0.5\,h^{-1}$Mpc.

\section{A New Model for the Halo Spin-Spin Correlations}

Now that the extended model for the tidally induced spin alignments are found to successfully work on the dwarf galactic scale, 
we would like to construct a new improved formula for the spin-spin correlations of abundant dwarf galactic halos in the framework of 
this model, in the hope that it could describe well the behavior of the halo spin-spin correlation function at large distances. 
But, before embarking on this task, we would like to briefly review the previous formulae that have paved a path to this new one. 

\subsection{Review of the Previous Models}\label{sec:review_eta}

\citet{pen-etal00} defined the halo spin-spin correlation function, $\eta(r)$, as an ensemble average of the square of the dot product of 
the unit spin vectors of two galactic halos at the positions of $\bx$ and $\bxp$, respectively, with separation distance  
$r\equiv\vert\bx-\bxp\vert$: 
\begin{equation}
\label{eqn:eta}
\eta(r) \equiv \langle\vert\hbs(\bx)\cdot\hbsp(\bxp)\vert^{2}\rangle - \frac{1}{3} \, .
\end{equation}
Note that $\eta(r)$ vanishes at large $r$ since the first ensemble average term in the right-hand side (RHS) reaches $1/3$ at large $r$. 

Putting Equation (\ref{eqn:model1}) into Equation (\ref{eqn:eta}), \citet{pen-etal00} derived the following approximate formula for $\eta(r) $ with the 
help of Wick's theorem under the assumption that the surrounding tidal field is Gaussian and isotropic \citep[see also Appendix H in][]{lp01}
\footnote{In \citet{lp01}, the spin correlation parameter $a$ equals $3c_{t}/5$}. 
\ben
\label{eqn:eta_lin0}
\eta(r) &=&  \frac{9}{25}c^{2}_{t}\langle\hT_{ik}\hTp_{il}\hT_{kj}\hTp_{lj}\rangle - \frac{1}{3} \\
\label{eqn:eta_lin1}
&\approx & \frac{9}{25}c^{2}_{t}\frac{\langle\tT_{ik}\tT_{il}\tTp_{kj}\tTp_{lj}\rangle}{\langle\vert\tT\vert^{2}\vert\tTp\vert^{2}\rangle} - \frac{1}{3} \\
\label{eqn:eta_lin}
&\approx & a^{2}_{\rm nl}\frac{\xi^{2}(r)}{\sigma^{2}}\, ,
\een
where  $\xi(r)$ and $\sig^{2}$ are the two-point and auto correlation functions of the linear density field smoothed on $R_{f}$, respectively. 

The proportionality constant factor, $a_{\rm nl}$, between $\eta(r)$ and the rescaled density correlation $\txi(r)\equiv \xi(r)/\sigma^{2}$ 
is close to $3c^{2}_{t}/50$ if the smoothing scale $R_{f}$ equals the Lagrangian radius of the halo mass, $R_{\rm th}$ \citep{lp01}. 
If $R_{f}$ is different from $R_{\rm th}$, the value of $a_{\rm nl}$ would differ from $3c^{2}_{t}/50$. \citet{pen-etal00} treated this 
proportionality constant factor, $a_{\rm nl}$, as an adjustable parameter, given all the uncertainties involved in the approximations  
made in the derivation of Equations (\ref{eqn:eta_lin1})-(\ref{eqn:eta_lin}) as well as the difference between $R_{f}$ and $R_{\rm th}$.  
The key prediction of Equation (\ref{eqn:eta_lin}) was that as $\eta(r)$ drops with $r$ as rapidly as $\xi^{2}(r)$, the halo spin-spin correlation 
signal should be negligibly small at distances larger than a few megaparsecs \citep{pen-etal00,lp01}, which was later contradicted by several 
numerical results \citep{HZ02,lp08,HZ08}.

What failed Equation (\ref{eqn:eta_lin}) was not only the disagreements with the numerical results but also the limitation of its critical assumption that 
the tidal fields are Gaussian and isotropic. Although this simple assumption allowed the halo spin-spin correlation function to be expressed in terms of 
the linear observables,  it was to fail for the description of the spin-spin correlation function of the galactic halos since the tidal field on the galactic scale 
is neither Gaussian nor isotropic.  
It was \citet{HZ02} who pointed out the unrealistic assumption that underlies Equation (\ref{eqn:eta_lin}) about the Gaussianity of the tidal fields and 
claimed that the growth of the non-Gaussianity would drive $\eta(r)$ to be proportional to $\txi(r)$ rather than $\txi^{2}(r)$, which was confirmed by 
the subsequent numerical work of \citet{lp08}. 

To take into account the effect of the non-Gaussian tidal fields on the halo spin-spin correlations, \citet{lp08} suggested the following 
simple modification of Equation (\ref{eqn:eta_lin}) in the hope that it would yield a better agreement with the numerical results:    
\beq
\label{eqn:eta_ng}
\eta(r) \approx  a_{l}\txi^{2}(r) + \epsilon_{\rm nl}\txi(r)\, ,
\eeq
where $a_{l}$ and $\epsilon_{\rm nl}$ are two adjustable parameters.  Determining the best-fit values of $a_{l}$ and $\epsilon_{nl}$ by fitting Equation 
(\ref{eqn:eta_ng}) to the numerical results obtained from the Millennium Run N-body simulations \citep{millennium05}, \citet{lp08} demonstrated that for the galactic 
halos with masses in the range of $1.72\le M/(10^{11}\,h^{-1}\,M_{\odot})\le 10^{2}$ at redshifts $z\le 0.5$, the second term proportional to $\txi(r)$ dominate the first 
term proportional to $\txi^{2}(r)$ in Equation (\ref{eqn:eta_ng}). 

\subsection{Modeling the Anisotropic Tidal Effect}\label{sec:eta_ani}

Although \citet{lp08} found the linear scaling model, Equation (\ref{eqn:eta_ng}), to work much better than the quadractic scaling model, Equation (\ref{eqn:eta_lin}), 
\citet{lp08} also detected a low but significant signal of the halo spin-spin correlation at distances as large as $r\ge 10\,h^{-1}$Mpc, which could not be 
described by Equation (\ref{eqn:eta_ng}) even with the best-fit parameters. The non-zero value of $\eta(r)$ at $r\ge 10\,h^{-1}\,{\rm Mpc}$ turned out to be 
more significant especially at lower redshifts, $z\le 0.5$, which implies that some additional nonlinear effect other than the non-Gaussianity must be responsible 
for the presence of this signal. To find an improved model that can describe the spin-spin correlations at $r\ge 10\,h^{-1}$Mpc, 
we employ the extended model for the tidally induced spin alignments reviewed in Section \ref{sec:review_etm}.

Putting Equation (\ref{eqn:model2}) into the definition of $\eta(r)$, i.e,. Equation (\ref{eqn:eta}), and using the same approximation made by \citet{pen-etal00}, 
we have
\ben
\label{eqn:eta_ani1}
\eta(r) &\approx&  \frac{9}{25}d^{2}_{t}\langle\hT_{ij}\hTp_{ij}\rangle - \frac{1}{3} \\
&\approx&  \frac{18}{25}d^{2}_{t}\frac{\langle\tT_{ij}\tTp_{ij}\rangle}{\langle\vert\tTp\vert^{2}\rangle} 
- \frac{1}{3}\, .
\label{eqn:eta_ani2}
\een
To see how the assumption of the isotropic tidal field leads $\langle\tT_{ij}\tT_{ij}\rangle$ to be expressed in terms of $\txi(r)$, 
let us recall the following expression of $\langle\tT_{ij}\tT_{kl}\rangle$ \citep{lp01}:  
\beq
\langle\tT_{ij}\tTp_{kl}\rangle=
\langle T_{ij}\pT_{kl}\rangle - \frac{1}{3}\delta_{kl}\langle T_{ij}\pT_{nn}\rangle -
\frac{1}{3}\delta_{ij}\langle T_{mm}\pT_{kl}\rangle +  \frac{1}{9}\delta_{ij}\delta_{kl}\langle T_{mm}\pT_{nn}\rangle\, ,
\label{eqn:til_tt_cov}
\eeq
where
\ben  
\langle T_{ij}T^{\prime}_{kl}\rangle &=& 
(\delta_{ij}\delta_{kl}+\delta_{ik}\delta_{jl}+\delta_{il}\delta_{jk})
\left[\frac{J_{3}(r)}{6} - \frac{J_{5}(r)}{10}\right] + 
\hat{r}_i\hat{r}_j\hat{r}_k\hat{r}_l\left[\xi(r) + \frac{5J_{3}(r)}{2} - \frac{7J_{5}(r)}{2}\right] \nonumber \\
&&+\left(\delta_{ij}\hat{r}_k\hat{r}_l + \delta_{ik}\hat{r}_j\hat{r}_l 
+ \delta_{il}\hat{r}_k\hat{r}_j + \delta_{jk}\hat{r}_i\hat{r}_{l}
+ \delta_{jl}\hat{r}_i\hat{r}_k + \delta_{kl}\hat{r}_i\hat{r}_j\right)\left[\frac{J_{5}(r)}{2}-\frac{J_{3}(r)}{2}\right],    
\label{eqn:tt_cov}
\een
with
\beq
\label{eqn:j3j5}
J_{3}(r) = \frac{3}{r^{3}}\int_0^r \xi(r^{\prime}) r^{\prime 2} dr^{\prime 2}\, , \quad
J_{5}(r) = \frac{5}{r^{5}}\int_0^r \xi(r^{\prime}) r^{\prime 4} dr^{\prime 4}\, .
\eeq

Equation (\ref{eqn:tt_cov}) holds true only if the tidal field is isotropic.  If $i=k$ and $j=l$, then all of the terms containing $J_{3}$ and $J_{5}$ 
in Equation (\ref{eqn:til_tt_cov}) would vanish by symmetry, resulting in $\langle \tT_{ij}\tTp_{ij}\rangle$  expressed only in terms of $\xi(r)$. 
In the nonlinear regime, however, $\bT$ is far from being isotropic, the manifestation of which is nothing but the presence of the filamentary cosmic web 
\citep{cosmicweb96}. Assuming here that for the anisotropic $\bT$, the terms containing  $J_{3}$ and $J_{5}$ in the expression of 
$\langle \tT_{ij}\tTp_{ij}\rangle$ would not vanish, we propose the following fitting formula for $\eta(r)$: 
\beq
\label{eqn:eta_ani}
\eta(r) \approx \frac{18}{25}d^{2}_{t}\left[\txi(r) + g_{3}\tJ_{3}(r) - g_{5}\tJ_{5}(r)\right]\, ,
\eeq
where $\tJ_{3}\equiv J_{3}/\sig$, $\tJ_{5}\equiv J_{5}/\sig$, while $g_{3}$ and $g_{5}$ are two adjustable parameters. As $\bT$ 
becomes more anisotropic, the second and third terms containing $\tJ_{3}$ and $\tJ_{5}$, respectively, would become more 
dominant over the first term containing $\txi(r)$ in Equation (\ref{eqn:eta_ani}). In other words, the values of $g_{3}$ and $g_{5}$ 
would increase as $M$ and $z$ decreases. 

\subsection{Numerical Tests}\label{sec:eta_test}

Now, we are going to numerically test the validity of Equations (\ref{eqn:eta_ani1})-(\ref{eqn:eta_ani}) using the same datasets from the $\nu^{2}$GC-H2 simulation 
that are described in Section \ref{sec:etm_extension}.  We first divide an interval $0\le r/(h^{-1}{\rm Mpc})\le 20$ into short bins of the same length, $\Delta r$.  
Then, we compute the ensemble average of $\langle\vert\hbs\cdot\hbsp\vert^{2}\rangle$ over all of the halo pairs whose values of $r$ fall in each bin to 
numerically obtain $\eta(r)$ as defined in Equation (\ref{eqn:eta}). The one standard deviation of $\langle\vert\hbs\cdot\hbsp\vert^{2}\rangle$ is also calculated 
as the associated errors. In a similar manner, we compute $\langle\hT_{ij}({\bx})\hTp_{ij}({\bx}+{\br})\rangle$ from the reconstructed tidal tensors and determine 
the mean value of $d_{t}$ by Equation (\ref{eqn:dt}) from the measured values of $\hbs$ and $\hlam$ for each halo.  
Multiplying $\langle\hT_{ij}\hTp_{ij}\rangle$ by $9d^{2}_{t}/25$ and subtracting $1/3$ from it, we numerically obtain the RHS of Equation (\ref{eqn:eta_ani1}). 

\begin{figure}
\begin{center}
\includegraphics[scale=1.0]{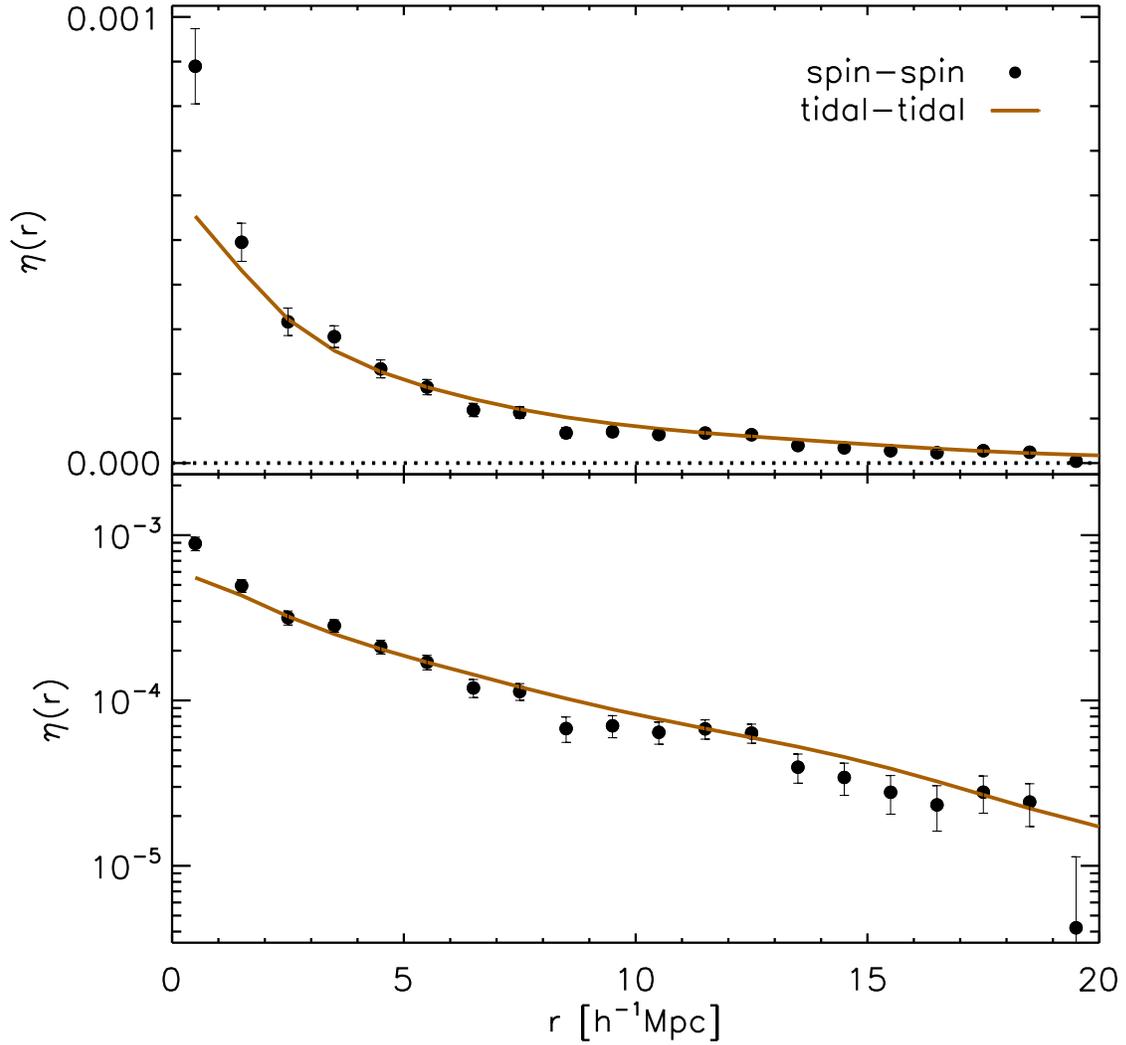}
\caption{Correlations of the unit spin vectors between the neighbor dwarf galactic halos (filled circles) and 
the rescaled correlations of the unit traceless tidal tensors (brown solid line) between their positions as a function 
of the separation distance at $z=0$ in the linear (top panel) and logarithmic scale (bottom panel).}
\label{fig:cross}
\end{center}
\end{figure}

Figure \ref{fig:cross} plots the numerically obtained $\eta(r)$ (filled circles) and compares it with the numerically obtained RHS of Equation (\ref{eqn:eta_ani1}) 
(brown sold lines) in the top panel. To show more clearly how $\eta(r)$ behaves at large distances, the bottom panel of Figure \ref{fig:cross} plots the same as 
the top panel but in the logarithmic scale. As can be seen, except for the disagreement at the first bin from the left which corresponds to $r\le 0.5\,h^{-1}$Mpc,  
the RHS of Equation (\ref{eqn:eta_ani1}) describes quite well the overall amplitude and behavior of $\eta(r)$. 
Regarding the disagreement at the first bin, we suspect that it  is likely caused by the resolution limit of the $\nu^{2}$GC-H2 simulation. 

\begin{figure}
\begin{center}
\includegraphics[scale=1.0]{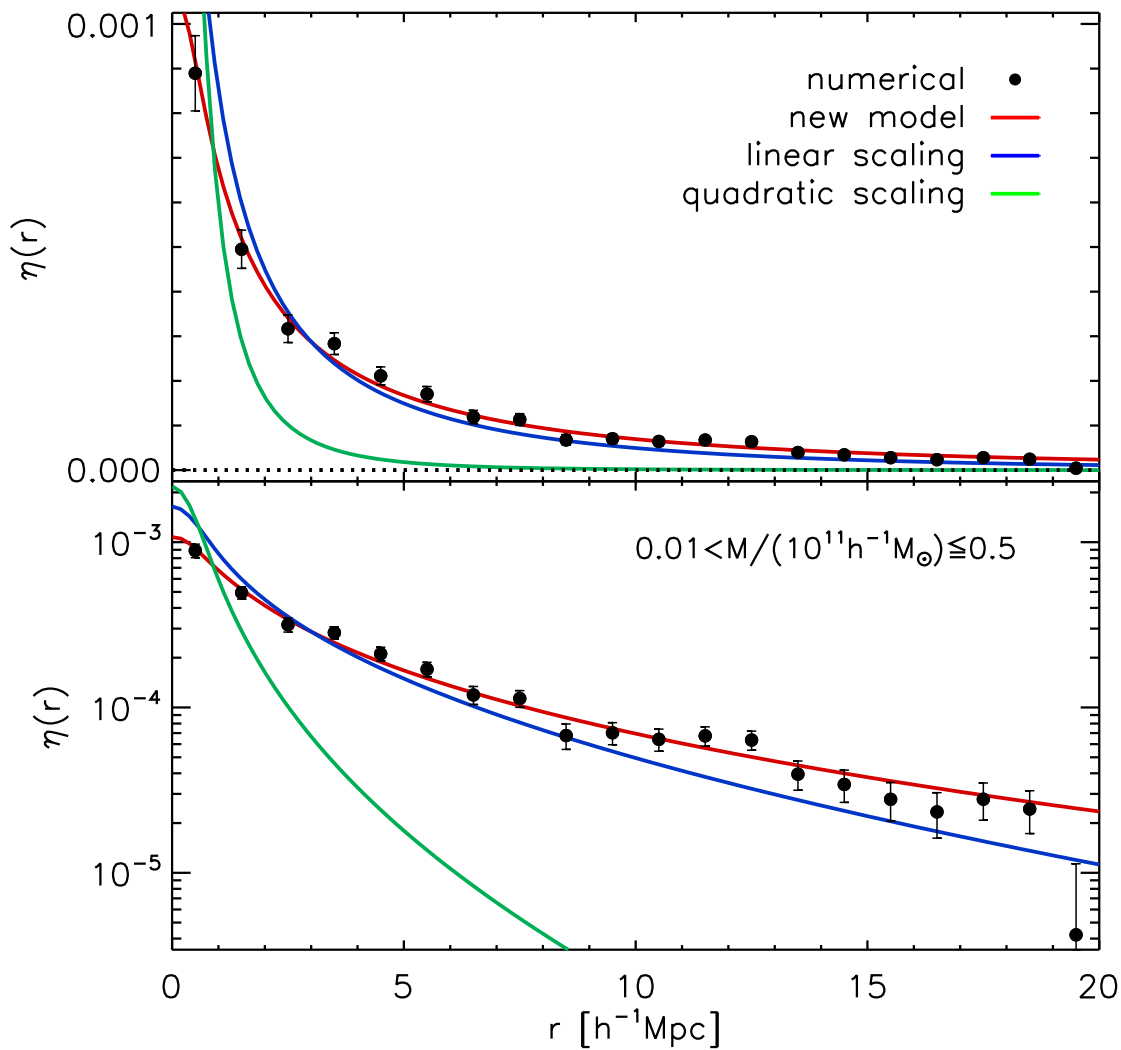}
\caption{Comparison of the numerically obtained spin-spin correlations of the dwarf galactic halos (filled circles) 
with three different formulae, Equations (\ref{eqn:eta_lin}), (\ref{eqn:eta_ng}) and (\ref{eqn:eta_ani}) with their best-fit 
parameters (green, blue and red solid lines, respectively.)}
\label{fig:cross_dwarf}
\end{center}
\end{figure}

Now that the validity of Equation (\ref{eqn:eta_ani1}) is confirmed, we would like to verify the usefulness of our new formula, Equations (\ref{eqn:eta_ani}).  
We determine the best-fit values of $g_{3}$ and $g_{5}$ by fitting Equation (\ref{eqn:eta_ani}) to the numerically obtained $\eta(r)$ with the help of the 
$\chi^{2}$-statistics (see Table \ref{tab:bestfit}). Figure \ref{fig:cross_dwarf} plots Equation (\ref{eqn:eta_ani}) with the best-fit parameters (red solid lines) and 
compares it with the numerical results (filled circles). 
We also fit Equations (\ref{eqn:eta_lin})-(\ref{eqn:eta_ng}) to the numerical results by adjusting $a_{\rm nl}$ and $\{a_{l}, \epsilon_{nl}\}$, respectively, 
which are shown in Figure \ref{fig:cross_dwarf} (green and blue solid lines, respectively). 
As can be seen, Equation (\ref{eqn:eta_lin}) that predicts the quadratic scaling of $\eta(r)$ with $\txi(r)$ grossly fails to describe $\eta(r)$ not only at 
large distances but everywhere.  Equation (\ref{eqn:eta_ng}) that predicts the linear scaling of $\eta(r)$ with $\txi(r)$ works better than Equation (\ref{eqn:eta_lin}), 
but fails to match the non-vanishing tail of $\eta(r)$ at $r\ge 10\,h^{-1}$Mpc, whose presence is believed to be induced by the anisotropic tidal effect.
Whereas, Equation (\ref{eqn:eta_ani}) matches the numerical results excellently in the whole range of $r$. 

To see whether or not Equation (\ref{eqn:eta_ani}) still works better than Equation (\ref{eqn:eta_ng}) even for the case of $R_{f}\ge 5\,h^{-1}$Mpc 
and $M\ge 5\times 10^{10}\,h^{-1}\,M_{\odot}$, we test it against the Small MultiDark Planck simulations (SMDPL) that has a larger simulation box of 
$L_{\rm box}=400\,h^{-1}$Mpc and a lower mass-resolution of $m_{p}=9.63\times 10^{7}\,h^{-1}\,M_{\odot}$ than the $\nu^{2}$GC-H2 simulation 
\citep{multidark}. Basically, we use the same dataset that \citet{lee19} compiled and used, which contains the sample of the galactic halos with 
$0.5\le M/(10^{11}\,h^{-1}\,M_{\odot})\le 50$ at $z=0$ and the values of $d_{t}$ measured at the location of each galactic halo from the alignments between 
the unit spin directions of the halos and the reconstructed local tidal field smoothed on the scale of $R_{f}=5\,h^{-1}$Mpc. 

\begin{figure}
\begin{center}
\includegraphics[scale=1.0]{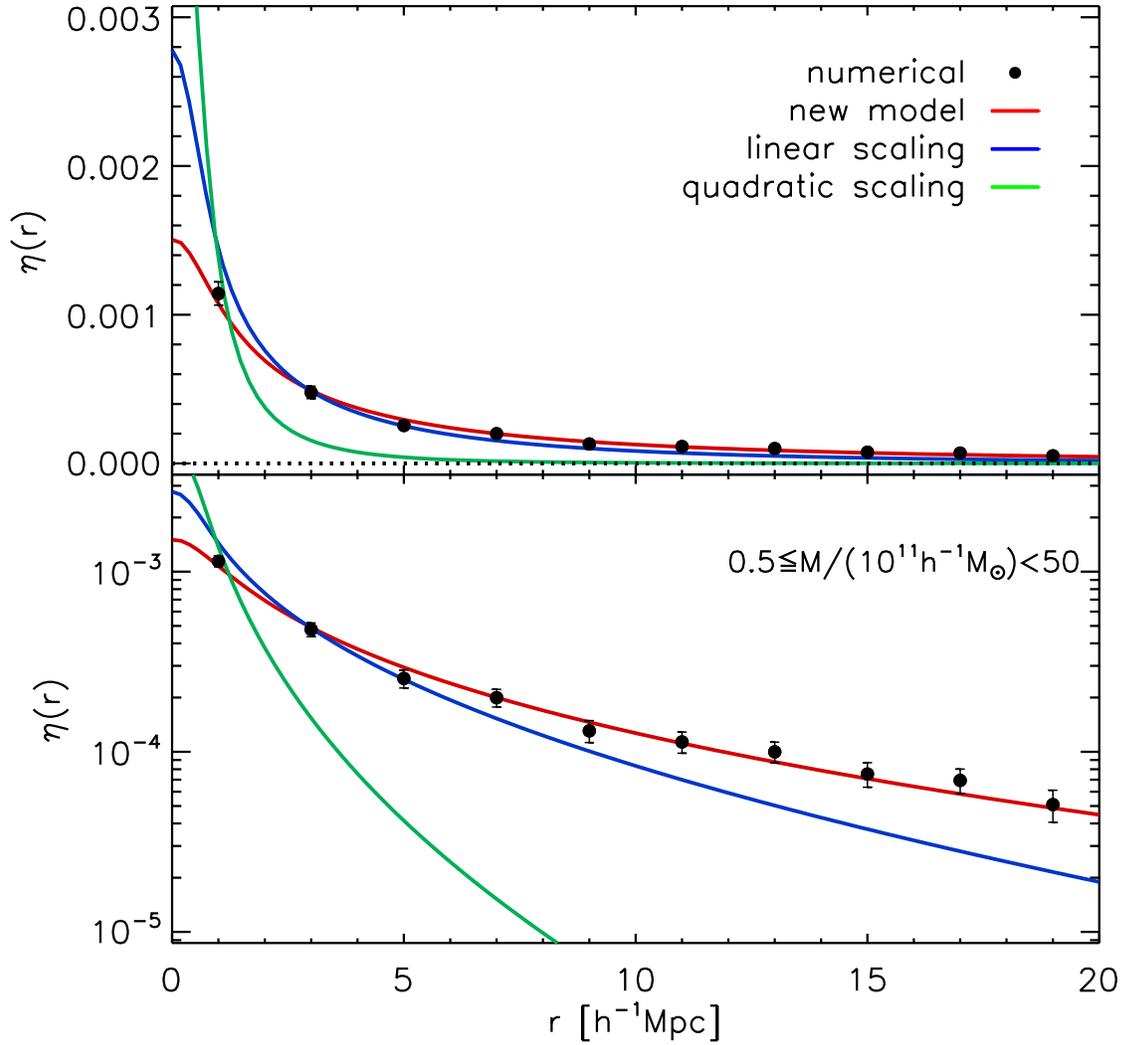}
\caption{Same as Figure \ref{fig:cross_dwarf} but for the higher-mass galactic halos in the 
range of $0.5\le M/(10^{11}\,h^{-1}\,M_{\odot}) \le 50$. }
\label{fig:cross_hmodel}
\end{center}
\end{figure}

Repeating the same analysis described in the above but with the halo sample from the SMDPL, we numerically determine $\eta(r)$, to which 
Equations (\ref{eqn:eta_lin}), (\ref{eqn:eta_ng}) and (\ref{eqn:eta_ani}) are fitted to find the best-fit values of their parameters (see Table \ref{tab:bestfit}). 
Regarding the value of $d_{t}$ in Equation (\ref{eqn:eta_ani}), we put the mean value averaged over the galactic halos in the sample from the SMDPL. 
Figure \ref{fig:cross_hmodel} plots the same as Figure \ref{fig:cross_dwarf} but with the sample from the SMDPL.  As can be seen, the new formula, 
Equation (\ref{eqn:eta_ani}), with the best-fit values of $\{g_{3}, g_{5}\}$ agrees best with the numerical results, even for the case of the 
higher mass halos and the much larger smoothing scale. 

\begin{figure}
\begin{center}
\includegraphics[scale=1.0]{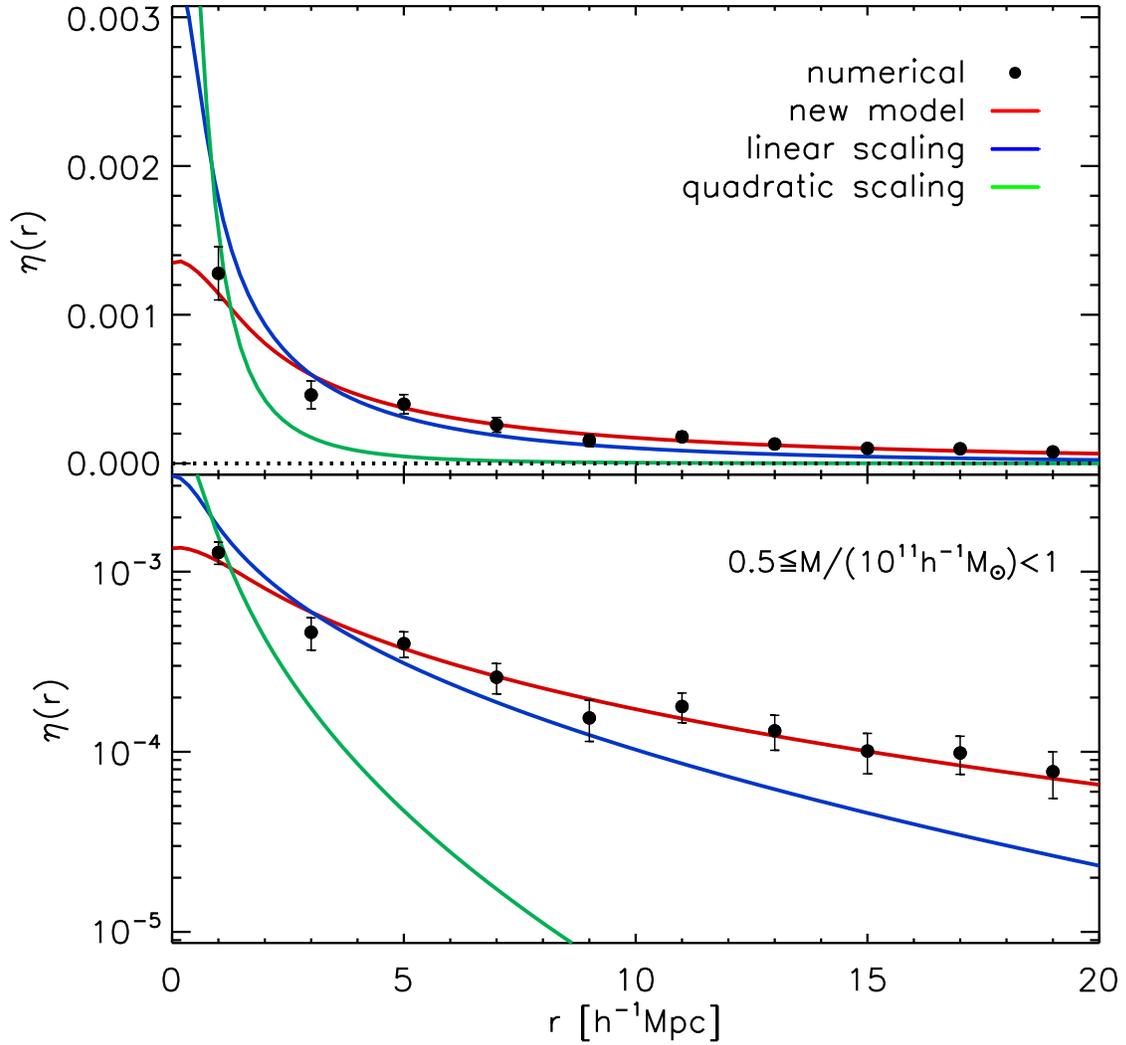}
\caption{Same as Figure \ref{fig:cross_hmodel} but for the galactic halos in the narrower mass range of 
$0.5\le M/(10^{11}\,h^{-1}\,M_{\odot}) \le 1$.} 
\label{fig:cross_hmodel1}
\end{center}
\end{figure}

\begin{figure}
\begin{center}
\includegraphics[scale=1.0]{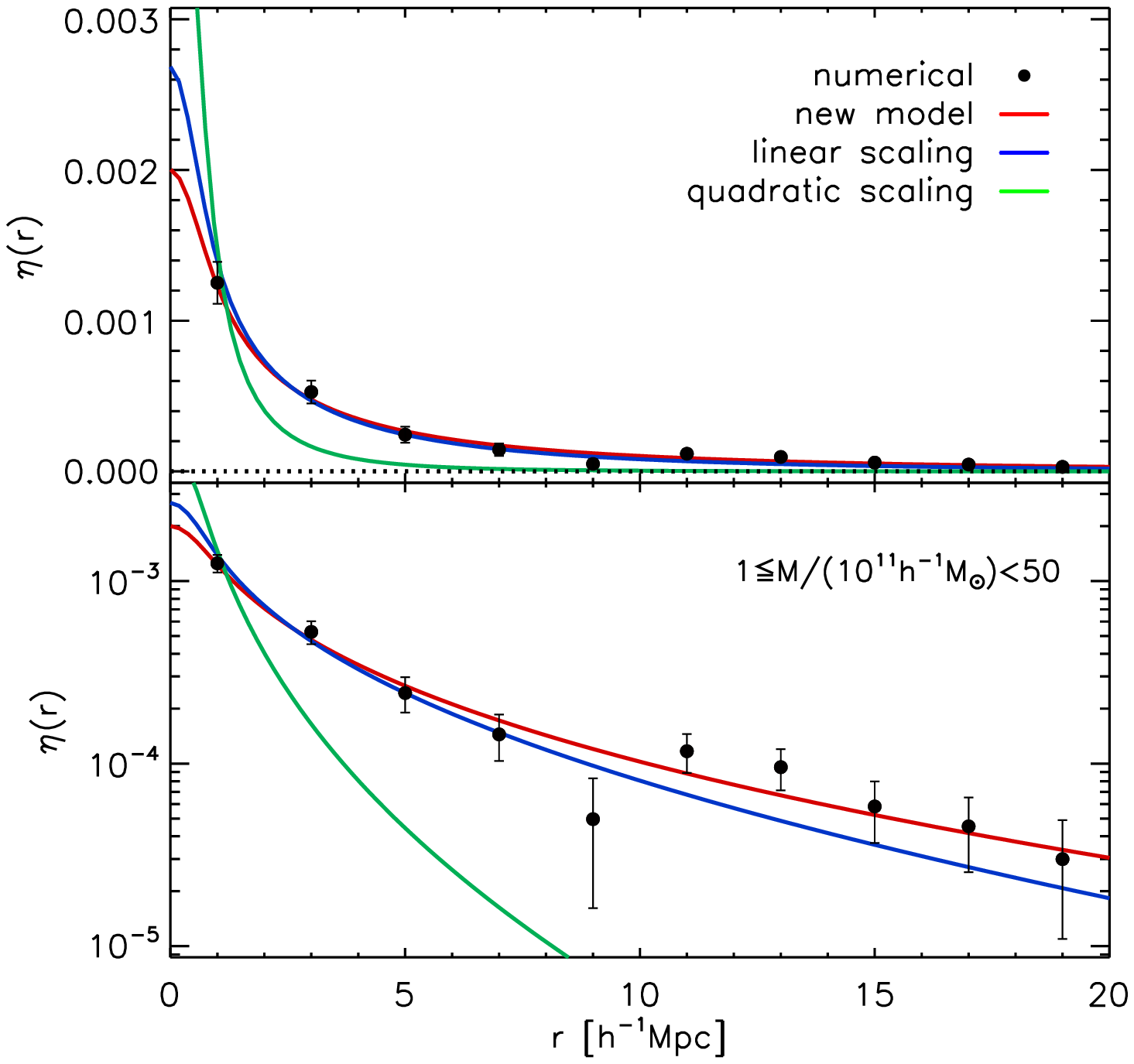}
\caption{Same as Figure \ref{fig:cross_hmodel1} but for the galactic halos in the higher mass range of 
$1\le M/(10^{11}\,h^{-1}\,M_{\odot}) \le 50$.}
\label{fig:cross_hmodel2}
\end{center}
\end{figure}

To see if the success of Equation (\ref{eqn:eta_ani}) depends on the halo mass, we split the halo sample from the SMDPL into two subsamples: 
One contains the halos in the mass range of $0.5\le M/(10^{11}\,h^{-1}\,M_{\odot}) \le 1$, while the rest of the halos with masses 
$1\le M/(10^{11}\,h^{-1}\,M_{\odot}) \le 50$ belong to the other subsample. We perform the same analysis but with each subsample separately, 
the results of which are displayed in Figures \ref{fig:cross_hmodel1}-\ref{fig:cross_hmodel2}. 
As can be seen, the best agreement is achieved by Equation (\ref{eqn:eta_ani}) for both of the subsamples. 
For the subsample with higher-mass halos, however, we find the difference between Equations (\ref{eqn:eta_ani}) and (\ref{eqn:eta_ng}) to 
substantially diminish. This mass dependence of the fitting results indicates that the anisotropic tidal effect is less strong for the higher-mass halos and 
that the linear scaling of $\eta(r)$ with $\txi(r)$ is a fairly good approximation for the case of the halos with $M\ge 10^{11}\,h^{-1}\,M_{\odot}$. 

\begin{figure}
\begin{center}
\includegraphics[scale=1.0]{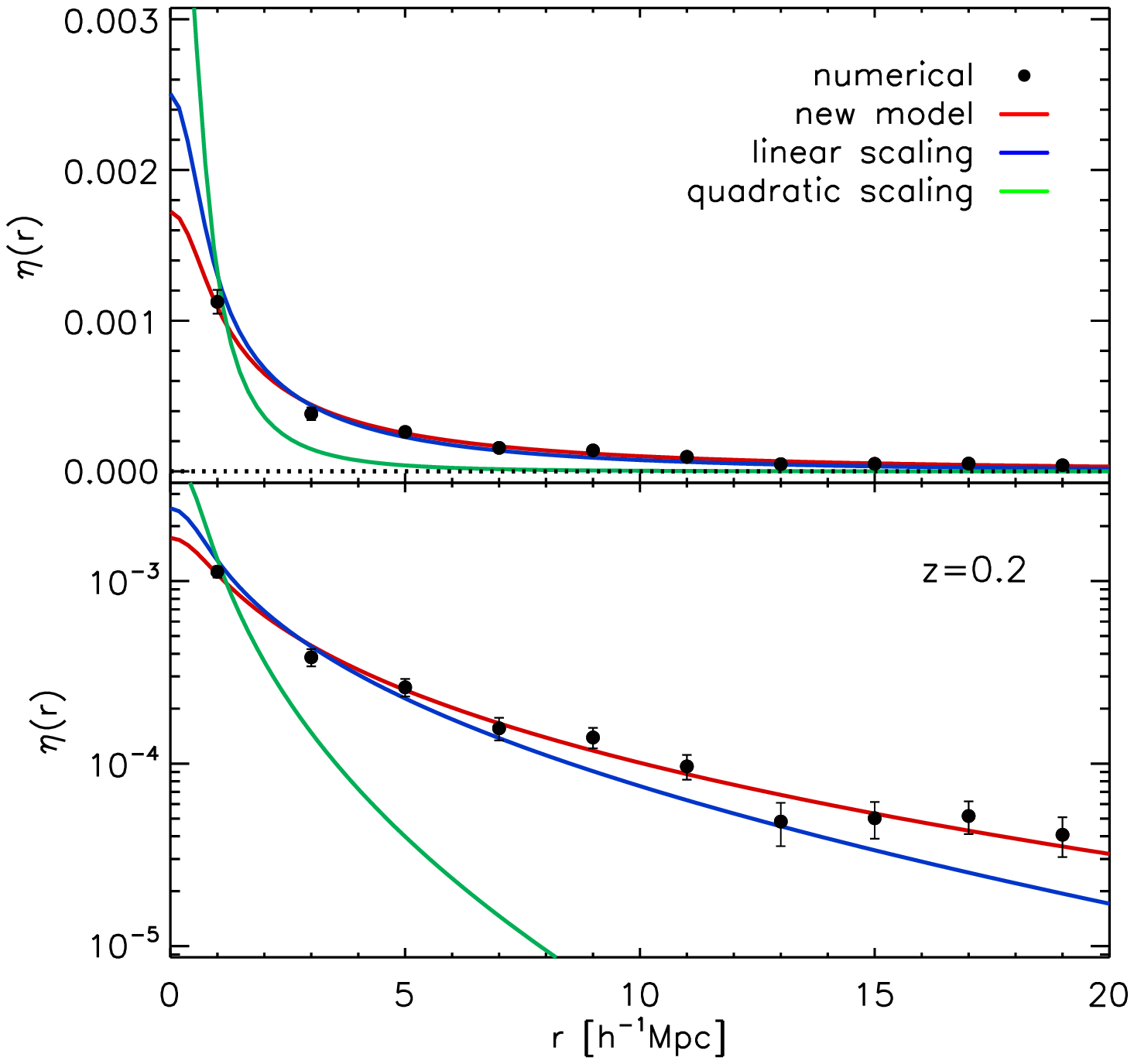}
\caption{Same as Figure \ref{fig:cross_hmodel} but at $z=0.2$.}
\label{fig:cross_z0.2}
\end{center}
\end{figure}

\begin{figure}
\begin{center}
\includegraphics[scale=1.0]{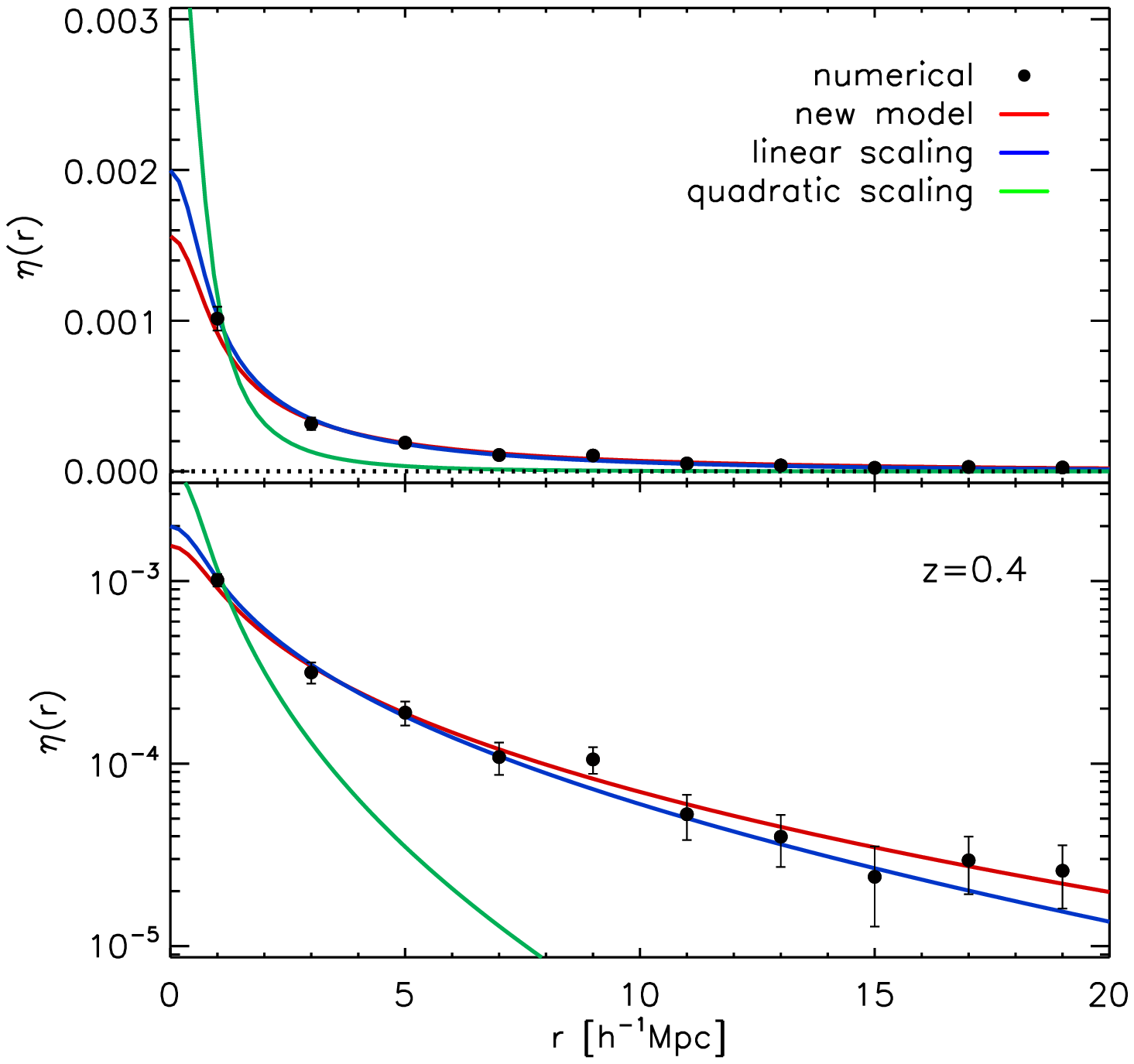}
\caption{Same as Figure \ref{fig:cross_hmodel} but at $z=0.4$.}
\label{fig:cross_z0.4}
\end{center}
\end{figure}

We also examine the validity of Equation (\ref{eqn:eta_ani}) at higher redshifts. Using the halos resolved at redshifts $0.2$ and $0.4$ from the 
SMDPL, we conducted the same analysis, the results of which are shown in Figures \ref{fig:cross_z0.2}-\ref{fig:cross_z0.4}, respectively. 
At both of the redshifts, Equation (\ref{eqn:eta_ani}) yields the best match to the numerically obtained $\eta(r)$. 
Note, however,  that as $\eta(r)$ drops relatively faster with $r$ at higher redshifts, the linear scaling, Equation (\ref{eqn:eta_ng}), provides a good 
match to the numerical results at $z=0.4$. Whereas, the original model suggested by \citet{pen-etal00} turned out to be invalid at both of the redshifts. 
The best-fit values of the two parameters, $g_{3}$ and $g_{5}$, of our new formula, Equation (\ref{eqn:eta_ani}), 
for various cases of $M$, $R_{f}$ and $z$ are listed in Table \ref{tab:bestfit}. As can be seen, the value of $g_{5}$ drops with the increment of $M$ and 
$z$ much more rapidly than $g_{3}$, which implies that the third term in Equation (\ref{eqn:eta_ani}) is the most sensitive indicator of the anisotropic 
tidal effect. 
 
\section{Summary and Discussion}\label{sec:sum}

Using the high-resolution N-body simulations, we have determined the spin-spin correlation function, $\eta(r)$, of DM halos in a broad mass range of 
$0.01\le M/(10^{11}\,h^{-1}\,M_{\odot})\le 50$ at $z=0$ and found it to decrease with the separation distance, $r$, much less rapidly than the rescaled 
two-point correlation function of the linear density field, $\txi(r)$, unlike the prediction of the previous model based on the tidal torque theory. 
However, the disagreement between the numerical results and the previous prediction has been shown to become smaller with the 
increment of $z$, almost vanishing at $z\ge 0.4$. 
Figuring out that the underlying assumption of the {\it isotropic} tidal field caused the disagreement, we have incorporated the anisotropic 
tidal effect into the previous model to derive a new formula with two fitting parameters for $\eta(r)$ expressed in terms of the integrals of $\txi(r)$.
The new formula with the best-fit parameters has turned out to excellently match the numerical results in the broad range of the halo masses, 
describing especially well the behavior of $\eta(r)$ at large distances of $r\ge 10\,h^{-1}$Mpc.  
 
Although our new model deals with the low-mass galactic halos in the highly nonlinear regime, it requires no higher order nor nonlinear statistics.  
The halo spin-spin correlations can still be linked by our new model to the linear observables, the integrals of the rescaled linear density two-point correlation 
functions, $\tJ_{3}(r)$ and $\tJ_{5}(r)$, which in turn implies that our model would allow us to reconstruct the rescaled linear density two-point correlation 
by measuring the spin-spin correlation of the galactic halos. Since the large-scale tail of $\txi(r)$ is sensitively dependent on the nature and amount of 
DM, our model for $\eta(r)$ at large distances of $r\ge 10\,h^{-1}$Mpc could be used as a complementary probe of DM. 
Furthermore, our model for $\eta(r)$ is independent of the amplitude of $\xi(r)$, it has a potential to break the degeneracy between the power spectrum 
amplitude and the amount of dark matter.  

To use our model for the halo spin-spin correlation function in practice as a probe of DM, however, the following issues must be addressed. 
The first issue is whether or not our formula for $\eta(r)$ is still valid even in the alternative non-$\Lambda$CDM cosmologies.  High-resolution N-body 
simulations performed for alternative cosmologies would be required to examine this and to investigate how the best-fit parameters of our model depend 
on the background cosmology. 
The second issue is whether or not our model can also validly describe the directly observable spin-spin correlation of the luminous galaxies.  
Hydrodynamic simulations showed that the spin directions of the baryonic parts are aligned not with those of the entire DM halos but 
rather with those  of the inner parts of the DM halos \citep[e.g.,][]{hah-etal10}. To apply our formula to describe the observed spin-spin correlation 
function of the luminous galaxies it will be necessary to examine whether or not the formula works well even when the spin directions are measured 
from the particles located in the inner central part of the DM halos. 

The third issue is how to measure the spin directions of the galaxies as accurately as possible from observations. For the cases of the giant late-type 
spiral galaxies whose position angles and axial ratios are available, the circular thin disk approximation has been conventionally employed to measure 
their spin directions \citep[e.g.,][]{LE07,lee11}. For the case of the elliptical and dwarf galaxies, however, the same approximation cannot be used for the 
measurements of their spin directions since their shapes obviously deviate far from a circular thin disk.   Detailed kinematic structures of the galaxies 
observed by a spectroscopic survey like MaNGA ("Mapping Nearby Galaxies at Apache Point Observatory") \citep{manga} may be useful to determine 
the three dimensional directions of the spin axes of those galaxies for which the conventional method fails (S. Kim private communication).

The fourth one is whether or not the same formula can be used to describe the galaxy spin-spin correlations measured in {\it redshift space} 
rather than in real space. In the original analysis of \citet{pen-etal00}, the linear density two-point correlation function was convolved with a Gaussian 
filter, $\exp\left[-r^{2}/(2\sigma^{2}_{r})\right]$, to derive the spin-spin correlation function in redshift space, where $\sigma_{r}\sim 1.3\,h^{-1}$Mpc 
is the typical velocity dispersion of a giant spiral galaxy \citep{dav-etal97}. However, since the galaxies with different types have different velocity dispersions, 
the errors associated with the values of $\sigma_{v}$ are likely to contaminate the weak signals of the galaxy spin-spin correlations at $r\ge 10\,h^{-1}$Mpc. 
A more elaborate method to recover the spin-spin correlation at large distances measured in redshift space will be necessary. Our future work is in the direction 
of resolving the above issues and testing the spin-spin correlation of the dwarf galaxies as a probe of DM. 

\acknowledgements

I gratefully acknowledge the website, {\tt www.skiesanduniverses.org}, for making the data from the $\nu^{2}$GC-H2 simulation 
accessible. The CosmoSim database used in this paper is a service by the Leibniz-Institute for Astrophysics Potsdam (AIP).
The MultiDark database was developed in cooperation with the Spanish MultiDark Consolider Project CSD2009-00064.î
I gratefully acknowledge the Gauss Centre for Supercomputing e.V. (www.gauss-centre.eu) and the Partnership for Advanced 
Supercomputing in Europe (PRACE, www.prace-ri.eu) for funding the MultiDark simulation project by providing computing time on 
the GCS Supercomputer SuperMUC at Leibniz Supercomputing Centre (LRZ, www.lrz.de).
The Bolshoi simulations have been performed within the Bolshoi project of the University of California High-Performance 
AstroComputing Center (UC-HiPACC) and were run at the NASA Ames Research Center.

I acknowledge the support of the Basic Science Research Program through the National Research Foundation (NRF) 
of Korea funded by the Ministry of Education (NO. 2016R1D1A1A09918491).  I was also partially supported by a research 
grant from the NRF of Korea to the Center for Galaxy Evolution Research (No.2017R1A5A1070354).

\clearpage
\clearpage
\clearpage
\clearpage
\clearpage
\clearpage
\clearpage
\clearpage
\begin{deluxetable}{ccccc}
\tablewidth{0pt}
\setlength{\tabcolsep}{5mm}
\tablecaption{Best-fit Parameters for the Halo Spin-Spin Correlations}
\tablehead{$z$ & $M$ & $R_{f}$ & $g_{3}$ & $g_{5}$ \\
& $(10^{11}\,h^{-1}M_{\odot})$ & ($h^{-1}\,$Mpc)& &}
\startdata
$0.0$  & $[0.01,\ 0.5)$   & $0.5$ & $2.1$ & $2.3$ \\
$0.0$  & $[0.5,\ 50.0)$ & $5.0$ & $6.0$ & $5.1$ \\
$0.0$ & $[0.5,\ 1.0)$ & $5.0$ & $8.3$ & $8.0$  \\
$0.0$  & $[1.0,\ 50.0)$ & $5.0$ & $2.4$ & $0.1$ \\
$0.2$ & $[0.5,\ 50.0)$ & $5.0$ & $3.0$ & $1.8$  \\
$0.4$  & $[0.5,\ 50.0)$ & $5.0$ & $1.0$ & $0.0$ 
\enddata
\label{tab:bestfit}
\end{deluxetable}


\clearpage
\begin{thebibliography}{000}
\bibitem[Arag{\'o}n-Calvo et al.(2007)]{ara-etal07} 
Arag{\'o}n-Calvo, M.~A., van de Weygaert, R., Jones, B.~J.~T., \& van der Hulst, J.~M.\ 2007, \apjl, 655, L5 
\bibitem[Aslanyan et al.(2016)]{asl-etal16} 
Aslanyan, G., Price, L.~C., Adams, J., et al.\ 2016, \prl, 117, 141102
\bibitem[Banik, \& Zhao(2018)]{BZ18} 
Banik, I., \& Zhao, H.\ 2018, \mnras, 480, 2660
\bibitem[Behroozi et al.(2013)]{rockstar} 
Behroozi, P.~S., Wechsler, R.~H., \& Wu, H.-Y.\ 2013, \apj, 762, 109
\bibitem[de Bernardis et al.(2000)]{boom00} 
de Bernardis, P., Ade, P.~A.~R., Bock, J.~J., et al.\ 2000, \nat, 404, 955
\bibitem[Bett et al.(2007)]{bet-etal07} 
Bett, P., Eke, V., Frenk, C.~S., et al.\ 2007, \mnras, 376, 215 
\bibitem[Bett \& Frenk(2012)]{BF12} 
Bett, P.~E., \& Frenk, C.~S.\ 2012, \mnras, 420, 3324 
\bibitem[Bett \& Frenk(2016)]{BF16} 
Bett, P.~E., \& Frenk, C.~S.\ 2016, \mnras, 461, 1338
\bibitem[Biagetti(2019)]{bia19} 
Biagetti, M.\ 2019, arXiv e-prints, arXiv:1906.12244
\bibitem[Bland-Hawthorn, \& Peebles(2006)]{BP06} 
Bland-Hawthorn, J., \& Peebles, P.~J.~E.\ 2006, Science, 313, 311
\bibitem[Bond et al.(1996)]{cosmicweb96} 
Bond, J.~R., Kofman, L., \& Pogosyan, D.\ 1996, \nat, 380, 603
\bibitem[Bundy et al.(2015)]{manga} 
Bundy, K., Bershady, M.~A., Law, D.~R., et al.\ 2015, \apj, 798, 7 
\bibitem[Carlesi et al.(2017)]{car-etal17} 
Carlesi, E., Mota, D.~F., \& Winther, H.~A.\ 2017, \mnras, 466, 4813
\bibitem[Chisari et al.(2016)]{chi-etal16} 
Chisari, N.~E., Dvorkin, C., Schmidt, F., et al.\ 2016, \prd, 94, 123507
\bibitem[Codis et al.(2012)]{cod-etal12} 
Codis, S., Pichon, C., Devriendt, J., et al.\ 2012, \mnras, 427, 3320
\bibitem[Codis et al.(2015a)]{cod-etal15a} 
Codis, S., Gavazzi, R., Dubois, Y., et al.\ 2015, \mnras, 448, 3391
\bibitem[Codis et al.(2015b)]{cod-etal15b} 
Codis, S., Pichon, C., \& Pogosyan, D.\ 2015, \mnras, 452, 3369
\bibitem[Crittenden et al.(2001)]{cri-etal01} 
Crittenden, R.~G., Natarajan, P., Pen, U.-L., \& Theuns, T.\ 2001, \apj, 559, 552
\bibitem[Davis et al.(1997)]{dav-etal97} 
Davis, M., Miller, A., \& White, S.~D.~M.\ 1997, \apj, 490, 63
\bibitem[Doroshkevich(1970)]{dor70} 
Doroshkevich, A.~G.\ 1970, Astrofizika, 6, 581
\bibitem[Ganeshaiah Veena et al.(2018)]{vee-etal18} 
Ganeshaiah Veena, P., Cautun, M., van de Weygaert, R., et al.\ 2018, \mnras, 481, 414
\bibitem[Hahn et al.(2007)]{hah-etal07} 
Hahn, O., Carollo, C.~M., Porciani, C., \& Dekel, A.\ 2007, \mnras, 381, 41 
\bibitem[Hahn et al.(2010)]{hah-etal10} 
Hahn, O., Teyssier, R., \& Carollo, C.~M.\ 2010, \mnras, 405, 274
\bibitem[Hui \& Zhang(2002)]{HZ02}
Hui, L. \& Zhang Z. 2002, preprint [astro-ph/0205512]
\bibitem[Hui, \& Zhang(2008)]{HZ08} 
Hui, L., \& Zhang, J.\ 2008, \apj, 688, 742
\bibitem[Ishiyama et al.(2015)]{ish-etal15} 
Ishiyama, T., Enoki, M., Kobayashi, M.~A.~R., et al.\ 2015, \pasj, 67, 61
\bibitem[Kogai et al.(2018)]{kog-etal18} 
Kogai, K., Matsubara, T., Nishizawa, A.~J., et al.\ 2018, JCAP, 2018, 014
\bibitem[Klypin et al.(2016)]{multidark} 
Klypin, A., Yepes, G., Gottl{\"o}ber, S., et al.\ 2016, \mnras, 457, 4340
\bibitem[Lacerna \& Padilla(2012)]{LP12} 
Lacerna, I., \& Padilla, N.\ 2012, \mnras, 426, L26 
\bibitem[Laigle et al.(2015)]{lai-etal15} 
Laigle, C., Pichon, C., Codis, S., et al.\ 2015, \mnras, 446, 2744
\bibitem[Lee \& Pen(2000)]{lp00} 
Lee, J., \& Pen, U.-L.\ 2000, \apjl, 532, L5
\bibitem[Lee \& Pen(2001)]{lp01} 
Lee, J., \& Pen, U.-L.\ 2001, \apj, 555, 106 
\bibitem[Lee \& Erdogdu(2007)]{LE07} 
Lee, J., \& Erdogdu, P.\ 2007, \apj, 671, 1248 
\bibitem[Lee \& Pen(2008)]{lp08}
Lee, J., \& Pen, U.-L.\ 2008, \apj, 681, 798
\bibitem[Lee(2011)]{lee11} 
Lee, J.\ 2011, \apj, 732, 99 
\bibitem[Lee(2019)]{lee19} 
Lee, J.\ 2019, \apj, 872, 37
\bibitem[Libeskind et al.(2013)]{lib-etal13} 
Libeskind, N.~I., Hoffman, Y., Forero-Romero, J., et al.\ 2013, \mnras, 428, 2489
\bibitem[Makiya et al.(2016)]{mak-etal16} 
Makiya, R., Enoki, M., Ishiyama, T., et al.\ 2016, \pasj, 68, 25
\bibitem[Ntampaka et al.(2017)]{nta-etal17} 
Ntampaka, M., Trac, H., Cisewski, J., et al.\ 2017, \apj, 835, 106
\bibitem[Park, \& Ratra(2018)]{PR18} 
Park, C.-G., \& Ratra, B.\ 2018, arXiv e-prints, arXiv:1801.00213
\bibitem[Paz et al.(2008)]{paz-etal08} 
Paz, D.~J., Stasyszyn, F., \& Padilla, N.~D.\ 2008, \mnras, 389, 1127 
\bibitem[Pen et al.(2000)]{pen-etal00} 
Pen, U.-L., Lee, J., \& Seljak, U.\ 2000, \apjl, 543, L107
\bibitem[Planck Collaboration et al.(2014)]{planck13} 
Planck Collaboration, Ade, P.~A.~R., Aghanim, N., et al.\ 2014, \aap, 571, A16
\bibitem[Porciani et al.(2002)]{por-etal02}
Porciani, C., Dekel, A., \& Hoffman, Y. 2002, \mnras, 332, 339
\bibitem[Schmidt et al.(2015)]{sch-etal15} 
Schmidt, F., Chisari, N.~E., \& Dvorkin, C.\ 2015, JCAP, 2015, 032
\bibitem[Seo, \& Eisenstein(2003)]{SE03} 
Seo, H.-J., \& Eisenstein, D.~J.\ 2003, \apj, 598, 720
\bibitem[Silva, \& Yunes(2019)]{SY19} 
Silva, H.~O., \& Yunes, N.\ 2019, arXiv e-prints, arXiv:1902.10269
\bibitem[Trowland et al.(2013)]{tro-etal13} 
Trowland, H.~E., Lewis, G.~F., \& Bland-Hawthorn, J.\ 2013, \apj, 762, 72 
\bibitem[Springel et al.(2005)]{millennium05} 
Springel, V., White, S.~D.~M., Jenkins, A., et al.\ 2005, \nat, 435, 629
\bibitem[Vittorio et al.(1986)]{vit-etal86} 
Vittorio, N., Juszkiewicz, R., \& Davis, M.\ 1986, \nat, 323, 132
\bibitem[Welker et al.(2014)]{wel-etal14} 
Welker, C., Devriendt, J., Dubois, Y., Pichon, C., \& Peirani, S.\ 2014, \mnras, 445, L46 
\bibitem[White(1984)]{whi84} 
White, S.~D.~M.\ 1984, \apj, 286, 38
\bibitem[Yu et al.(2019)]{yu-etal19} 
Yu, H.-R., Yu, Y., Motloch, P., et al.\ 2019, arXiv e-prints, arXiv:1904.01029
\bibitem[Zel'dovich(1970)]{zel70} 
Zel'dovich, Y.~B.\ 1970, \aap, 5, 84 
\bibitem[Zhang, Yang \& Faltenbacher(2009)]{zha-etal09} 
Zhang, Y., Yang, X., Faltenbacher, A., et al.\ 2009, \apj, 706, 747  
\end{thebibliography}
\end{document}